\documentclass[12pt, onecolumn]{IEEEtran}
\IEEEoverridecommandlockouts
\usepackage{graphicx}
\usepackage{subfigure}
\usepackage{color}
\usepackage{epsfig}
\usepackage{amssymb}
\usepackage{amsmath}
\usepackage{amsthm}
\usepackage{latexsym}

\usepackage{setspace}
\newcommand{\bPr}[1]{{\Pr}\left(#1\right)}

\newcommand{\bP}[2]{{P}_{#1}\left({#2}\right)}

\newcommand{\mcf}{\mathtt{mcf}}
\newcommand{\cX}{{\mathcal X}}
\newcommand{\cJ}{{\mathcal J}}
\newcommand{\cD}{{\mathcal D}}
\newcommand{\cY}{{\mathcal Y}}
\newcommand{\cZ}{{\mathcal Z}}

\newcommand{\cU}{{\mathcal U}}
\newcommand{\cM}{{\mathcal M}}
\newcommand{\cA}{{\mathcal A}}
\newcommand{\cK}{{\mathcal K}}
\newcommand{\cP}{{\mathcal P}}

\newcommand{\cV}{{\mathcal V}}
\newcommand{\cQ}{{\mathcal Q}}
\newcommand{\cR}{{\mathcal R}}

\newcommand{\ep}{\epsilon}
\newcommand{\la}{\lambda}

\newcommand{\cDn}{\cD^{(n)}}
\newcommand{\cRn}{\cR^{(n)}}
\newcommand{\tDn}{\tilde{\cD}^{(n)}}

\newtheorem{theorem}{Theorem}
\newtheorem{proposition}[theorem]{Proposition}

\newtheorem{lemma}[theorem]{Lemma}
\newtheorem*{lemmaGK}{Lemma GK}

\theoremstyle{remark}
\newtheorem*{remark*}{Remark}
\newtheorem*{remarks*}{Remarks}

\newtheorem{example}{Example}
\theoremstyle{definition}
\newtheorem{definition}{Definition}

\begin{document}
\IEEEoverridecommandlockouts

\title{When is a Function Securely Computable?}

\author{
\IEEEauthorblockN{Himanshu Tyagi, Prakash Narayan and  Piyush
Gupta} \footnote{The work of H. Tyagi and P. Narayan was supported
by the U.S. National Science Foundation under Grants CCF0635271
and CCF0830697. P. Gupta acknowledges support from NSF Grant
CNS-519535.

H. Tyagi and P. Narayan are with the Department of Electrical and
Computer Engineering and the Institute for Systems Research,
University of Maryland, College Park, MD 20742, USA. 
E-mail: {tyagi, prakash}@umd.edu

P. Gupta is with Bell Labs, Alcatel-Lucent, Murray Hill, NJ 07974,
USA. E-mail: pgupta@research.bell-labs.com }}

\maketitle

\begin{abstract}
A subset of a set of terminals that observe correlated signals
seek to compute a given function of the signals using public
communication. It is required that the value of the function be
kept secret from an eavesdropper with access to the communication.
We show that the function is securely computable if and only if
its entropy is less than the ``aided secret key" capacity of an
associated secrecy generation model, for which a single-letter
characterization is provided.
\end{abstract}


\section{Introduction}\label{s_int}
In an online auction, $m-1$ bidders acting independently of each
other, randomly place one of $k$ bids on a secure server. After a
period of independent daily bidding, the server posts a cryptic
message on a public website. Our results show that for $m > k+1$,
such a message exists from which each bidder can deduce securely
the highest bids, but no message exists to allow any of them to
identify securely the winners.

\begin{figure}
\vspace{.5cm}
\begin{center}
\epsfig{file=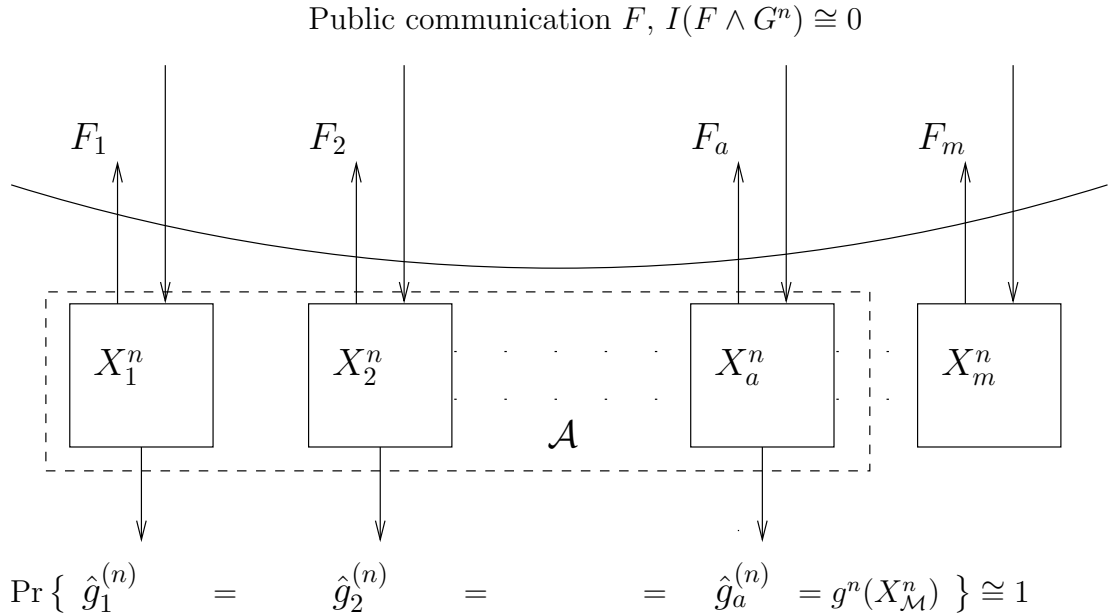, scale=1} \caption{Secure computation of $g$}
\label{f_SC}
\end{center}
\end{figure}

In general, suppose that the terminals in $\cM = \{1, \ldots, m\}$
observe correlated signals, and that a subset $\cA = \{1, \ldots,
a\}$ of them are required to compute ``securely" a given
(single-letter) function $g$ of all the signals. To this end,
following their observations, all the terminals are allowed to
communicate interactively over a public noiseless channel of
unlimited capacity, with all such communication being observed by
all the terminals. The terminals in $\cA$ seek to compute $g$ in
such a manner as to keep its value information theoretically
secret from an eavesdropper with access to the public
interterminal communication. See Figure \ref{f_SC}. A typical
application arises in a wireless network of colocated sensors
which seek to compute a given function of their correlated
measurements using public communication that does not give away
the value of the function.

Our goal is to characterize necessary and sufficient conditions
under which such secure computation is feasible. We formulate a
new Shannon theoretic multiterminal source model that addresses
the elemental question: {\emph{When can a function}} $g$ {\emph{be
computed so that its value is independent of the public
communication used in its computation}}?

We establish that the answer to this question is innately
connected to a problem of secret key (SK) generation in which all
the terminals in $\cM$ seek to generate ``secret common
randomness" at the largest rate possible, when the terminals in
$\cA^c = \cM/\cA$ are provided with side information for limited
use, by means of public communication from which an eavesdropper
can glean only a negligible amount of information about the SK.
The public communication from a terminal can be any function of
its own observed signal and of all previous communication. Side
information is provided to the terminals in $\cA^c$ in the form of
the value of $g$, and can be used only for recovering the key.
Such a key, termed an aided secret key (ASK), constitutes a
modification of the original notion of a SK in \cite{Mau93,
AhlCsi93, CsiNar04, CsiNar08}. The largest rate of such an ASK,
which can be used for encrypted communication, is the ASK capacity
$C$. Since a securely computable function $g$ for $\cA$ will yield
an ASK (for $\cM$) of rate equal to its entropy $H$, it is clear
that $g$ necessarily must satisfy $H\leq C$. We show that
surprisingly, $H< C$ is a sufficient condition for the existence
of a protocol for the secure computation of $g$ for $\cA$. When
all the terminals in $\cM$ seek to compute $g$ securely, the
corresponding ASK capacity reduces to the standard SK capacity for
$\cM$ \cite{CsiNar04, CsiNar08}. We also show that a function that
is securely computed by $\cA$ can be augmented by residual secret
common randomness to yield a SK for $\cA$ of optimum rate.

We also present the capacity for a general ASK model involving
\emph{arbitrary} side information at the secrecy-seeking set of
terminals for key recovery alone. Its capacity is characterized in
terms of the classic concept of ``maximum common function"
\cite{GacKor73}. Although this result is not needed in full dose
for characterizing secure computability, it remains of independent
interest.

We do not tackle the difficult problem of determining the minimum
rate of public communication needed for the secure computation of
$g$, which remains open even in the absence of a secrecy
constraint \cite{KorMar79}. Nor do we fashion efficient protocols
for this purpose. Instead, our mere objective in this work is to
find conditions for the {\emph{existence}} of such protocols.

The study of problems of function computation, with and without
secrecy requirements, has a long and varied history to which we
can make only a skimpy allusion here. Examples include: algorithms
for exact function computation by multiple parties (cf. e.g.,
\cite{Yao79,Gal88,GriKum05}); algorithms for asymptotically
accurate (in observation length) function computation (cf. e.g.,
\cite{OrlRoc01, MaIswGup09}); exact function computation with
secrecy (cf. e.g., \cite{OrlEl84}); and problems of oblivious
transfer \cite{NasWin06,AhlCsi07}.

Our results in Section \ref{s_res} are organized in three parts:
capacity of ASK model; characterization of the secure
computability of $g$; and a decomposition result for the total
entropy of the model. Proofs are provided in Section \ref{s_pro}
and concluding remarks in Section \ref{s_dis}.

\section{Preliminaries}\label{s_preliminaries}\label{s_pre}
Let $X_1, \ldots, X_m$, $m \geq 2$, be rvs with finite alphabets
$\cX_1, \ldots, \cX_m$, respectively. For any nonempty set $A
\subseteq \cM = \{1, \ldots, m\}$, we denote $X_A = (X_i,\ i \in
A)$. Similarly, for real numbers $R_1,\ldots, R_m$ and $A
\subseteq \cM$, we denote $R_A = (R_i,\ i \in A)$. Let $A^c$ be
the set $\cM\backslash A$. We denote $n$ i.i.d. repetitions of
$X_{\cM} = (X_1, \ldots, X_m)$ with values in $\cX_\cM = \cX_1
\times \ldots \times \cX_m$ by $X_{\cM}^n = (X_1^n, \ldots,
X_m^n)$ with values in $\cX_\cM^n = \cX_1^n\times \ldots \times
\cX_m^n$. Following \cite{CsiNar04}, given $\epsilon > 0$, for rvs
$U, V,$ we say that $U$ is $\epsilon$-\emph{recoverable} from $V$
if $\bPr{ U \neq f(V)} \leq \epsilon$ for some function $f(V)$ of
$V$. All logarithms and exponentials are with respect to the base
$2$.

We consider a multiterminal source model for secure computation
with public communication; this basic model was introduced in
\cite{CsiNar04} in the context of SK generation with public
transaction. Terminals $1, \dots, m$ observe, respectively, the
sequences $X_1^n, \ldots, X_m^n$, of length $n$. Let $g:\cX_\cM
\rightarrow \cY$ be a given mapping, where $\cY$ is a finite
alphabet. For $n \geq 1$, the mapping $g^n : \cX_\cM^n \rightarrow
\cY^n$ is defined by
\begin{align}\nonumber
g^n(x_\cM^n) &=(g(x_{11}, \ldots, x_{m1}), \ldots, g(x_{1n},
\ldots, x_{mn})), \\ \nonumber x_\cM^n &= (x_1^n, \ldots, x_m^n)
\in \cX_\cM^n.
\end{align}
For convenience, we shall denote the rv $g^n\left(X_\cM^n\right)$
by $G^n, n \geq 1$, and, in particular, $G^1 =
g\left(X_\cM\right)$ simply by $G$. The terminals in a given set
$\cA \subseteq \cM$ wish to ``compute securely" the function
$g^n(x^n_\cM)$ for $x^n_\cM$ in $\cX^n_\cM$. To this end, the
terminals are allowed to communicate over a noiseless public
channel, possibly interactively in several rounds. Randomization
at the terminals is permitted; we assume that terminal $i$
generates a rv $U_i,\ i \in \cM$, such that $U_1, \ldots, U_m$ and
$X_\cM^n$ are mutually  independent. While the cardinalities of range spaces
of $U_i, i \in \cM,$ are unrestricted, we assume that $H\left(U_\cM\right) < \infty$.
\begin{definition}\label{d_PubComm}
Assume without any loss of generality that the communication of
the terminals in $\cM$ occurs in consecutive time slots in $r$
rounds; such communication is described in terms of the mappings
\begin{align}\nonumber
f_{11},\ldots,f_{1m},
f_{21},\ldots,f_{2m},\ldots,f_{r1},\ldots,f_{rm},
\end{align}
with $f_{ji}$ corresponding to a message in time slot $j$ by
terminal $i$, $1 \leq j \leq r$, $1 \leq i \leq m$; in general,
$f_{ji}$ is allowed to yield any function of $(U_i, X_i^n)$ and of
previous communication described in terms of $\{ f_{kl}: k<j,\ l
\in \cM \ \text{or} \ k = j,\  l< i\}$. The corresponding rvs
representing the communication will be depicted collectively as
\begin{align}\nonumber
\mathbf{F} = \{F_{11},\ldots,F_{1m},
F_{21},\ldots,F_{2m},\ldots,F_{r1},\ldots,F_{rm}\},
\end{align}
where $\mathbf{F} = \mathbf{F}^{(n)}(U_\cM, X^n_\cM)$. A special
form of such communication will be termed \emph{noninteractive
communication} if $\mathbf{F} = \left(F_1, ..., F_m\right)$, where
$F_i = f_i\left(X_i^n\right)$, $i \in \cM$.
\end{definition}
\begin{definition}\label{d_SC1}
For $\ep_n > 0, n\geq 1$, we say that $g$ is
$\ep_n$-\emph{securely computable} ($\ep_n$- SC) by (the terminals
in) a given set $\cA \subseteq \cM$ with $|\cA| \geq 1$ from
observations of length $n$, randomization $U_\cM$ and public
communication $\mathbf{F} = \mathbf{F}^{(n)}$, if \vspace{0.1cm}
\\
\noindent(i) $g^n$ is $\ep_n$- recoverable from $(U_i, X_i^n,
\mathbf{F})$ for every $i \in \cA$, i.e., there exists
$\widehat{g}_i^{(n)}$ satisfying
\begin{align}\label{e_rel}
\bPr{\widehat{g}_i^{(n)}(U_i, X_i^n, \mathbf{F}) \neq G^n } \leq
\epsilon_n, \quad i \in \cA,
\end{align}
and \vspace{0.1cm}
\\
(ii) $g^n$ satisfies the ``strong" secrecy condition\footnote{The
notion of strong secrecy for SK generation was introduced in
\cite{Mau94}, and developed further in \cite{Csi96, CsiNar00}.}
\begin{align}\label{e_sec}
I(G^n \wedge \mathbf{F}) \leq \epsilon_n.
\end{align}
\end{definition}

By definition, an $\epsilon_n$-SC function $g$ is recoverable (as
$g^n$) at the terminals in $\cA$ and is effectively concealed from
an eavesdropper with access to the public communication
$\mathbf{F}$.
\begin{definition}\label{d_SC2}
We say that $g$ is \emph{securely computable} by $\cA$ if $g$ is
$\epsilon_n$- SC by $\cA$ from observations of length $n$,
suitable randomization $U_\cM$ and public communication
$\mathbf{F}$, such that $\displaystyle \lim_n \ep_n = 0$.
\end{definition}


\section{When is $g$ securely computable?}\label{s_res}
We consider first the case when all the terminals in $\cM$ wish to
compute securely the function $g$, i.e., $\cA = \cM$. Our result
for this case will be seen to be linked inherently to the standard
concept of  SK capacity for a multiterminal source model
\cite{CsiNar04, CsiNar08}, and serves to motivate our approach to
the general case when $\cA \subseteq \cM$.
\begin{definition}\label{d_SK} {\cite{CsiNar04,CsiNar08}}
For $\ep_n > 0, n\geq 1$, a function $K$ of $(U_\cM,X_\cM^n)$ is
an $\ep_n$-\emph{secret key} ($\ep_n$-SK) for (the terminals in) a
given set\footnote{For reasons of notation that will be apparent
later, we distinguish between the secrecy seeking set $\cA'
\subseteq \cM$ and the set $\cA\subseteq \cM$ pursuing secure
computation.} $\cA' \subseteq \cM$ with $|\cA'| \geq 2$,
achievable from observations of length $n$, randomization $U_\cM$
and public communication $\mathbf{F} = \mathbf{F}^{(n)}(U_\cM,
X^n_\cM)$ as above if \vspace{0.1cm}
\\(i) $K$ is $\ep_n$-recoverable from $(U_i, X_i^n, \mathbf{F})$ for every $i \in \cA'$;
\vspace{0.1cm}
\\(ii) $K$ satisfies the ``strong" secrecy condition
\begin{align}\label{e_sin}
\log|\cK| - H(K \mid \mathbf{F}) = \log|\cK| - H(K) + I(K \wedge
\mathbf{F}) \leq \ep_n,
\end{align}
where $\cK = \cK^{(n)}$ denotes the set of possible values of $K$.
The SK capacity $C(\cA')$ for $\cA'$ is the largest rate
$\displaystyle \lim_n \ (1/n)  \log |\cK^{(n)}|$ of $\ep_n$-SKs
for $\cA'$ as above, such that $\displaystyle \lim_n \ep_n = 0 $.
\end{definition}

\begin{remarks*}(i)
The secrecy condition (\ref{e_sin}) is tantamount jointly to a
nearly uniform distribution for $K$ (i.e., $\log|\cK| - H(K)$ is
small) and to the near independence of $K$ and $\mathbf{F}$ (i.e.,
$I(K \wedge \mathbf{F})$ is small).
\\(ii) For the trivial case $|\cA'| = 1$, clearly $C(\cA') = H(X_{\cA'})$.
\end{remarks*}

A single-letter characterization of the SK capacity $C(\cA')$ is
provided in \cite{CsiNar04,CsiNar08}.

\begin{theorem}\label{t_CSK}\emph{\cite{CsiNar04,CsiNar08}}
The SK capacity $C(\cA')$ equals
\begin{align}\label{e_CS}
C(\cA') = H(X_\cM) - R_{CO}(\cA'),
\end{align}
where
\begin{align}\label{e_RO}
R_{CO}(\cA') = \displaystyle \min_{R_\cM \in
\cR(\cA')}\sum_{i=1}^m R_i
\end{align}
with
\begin{align}\label{e_RCO}
\cR(\cA') = \bigg\{ R_\cM :  R_B \geq H(X_B \mid X_{B^c}),\quad B
\varsubsetneq \cM, \cA' \nsubseteq B\bigg\}.
\end{align}
Furthermore, the SK capacity can be achieved with noninteractive
communication and without recourse to randomization at the
terminals in $\cM$.
\end{theorem}

\begin{remark*}
The  SK capacity $C(\cA')$ is not increased if the secrecy
condition (\ref{e_sin}) is replaced by either of the following weaker requirements
\footnote{When randomization at the terminals in $\cM$ is not
permitted, the converse proof in \cite{CsiNar04} uses only the
first part of (\ref{e_sin'}) or (\ref{e_sin''}). When
randomization is allowed, since the cardinality of the range space
of $U_\cM$ is unrestricted, the converse proof in \cite{CsiNar04}
uses additionally the second part of (\ref{e_sin'}) or
(\ref{e_sin''}).}
\cite{Mau93,CsiNar04}:
\begin{align}\label{e_sin'}
\frac{1}{n} I(K \wedge \mathbf{F}) \leq \ep_n\quad \text{ and }\quad \frac{1}{n} \left(\log|\cK| -H(K) \right) \leq \ep_n,
\end{align}
or
\begin{align}\label{e_sin''}
\frac{1}{n} I(K \wedge \mathbf{F}) \leq \ep_n\quad \text{ and }\quad \limsup_n\frac{1}{n}\log|\cK| < \infty.
\end{align}

\end{remark*}

We recall from \cite{CsiNar04} that $R_{CO}(\cA')$ has the
operational significance of being the smallest rate of
``communication for omniscience" for $\cA'$, namely the smallest
rate $\displaystyle \lim_n \ (1/ n) \log \|\mathbf{F}^{(n)}\|$ of
suitable communication for the terminals in $\cM$ whereby
$X_\cM^n$ is $\ep_n$-recoverable from $(U_i , X_i^n,
\mathbf{F}^n)$ at each terminal $i \in \cA'$, with $\displaystyle
\lim_n  \ep_n = 0$; here $\|\mathbf{F}^{(n)}\|$ denotes the
cardinality of the set of values of $\mathbf{F}^{(n)}$. Thus,
$R_{CO}(\cA')$ is the smallest rate of interterminal communication
among the terminals in $\cM$ that enables every terminal in $\cA'$
to reconstruct with high probability all the sequences observed by
all the other terminals in $\cM$ with the cooperation of the
terminals in $\cM/\cA'$. The resulting omniscience for $\cA'$
corresponds to total ``common randomness" of rate $H(X_\cM)$. The
notion of omniscience, which plays a central role in SK generation
for the multiterminal source model \cite{CsiNar04}, will play a
material role in the secure computation of $g$ as well.

Noting that $g^n : \cX^n_\cM \rightarrow \cY^n$ implies
\begin{align}\label{e_cardboundg}
\frac{1}{n}\log \left| g^n\left(\cX^n_\cM\right)\right| \leq \log |\cX_\cM|,
\end{align}
a comparison of the conditions in (\ref{e_sec},
\ref{e_cardboundg}) and (\ref{e_sin''}) that must be met by a
securely computable $g$ and a SK $K$, respectively, shows for a
given $g$ to be securely computable, it is necessary that
\begin{align}\label{e_SCn}
H(G) \leq C(\cM).
\end{align}
Remarkably, it transpires that $H(G) < C(\cM)$ is a sufficient
condition for $g$ to be securely computable, and constitutes our
first result.

\begin{theorem}\label{t_SC}
A function $g$ is securely computable by $\cM$ if
\begin{align}\label{e_SCs}
H(G) < C(\cM).
\end{align}
Conversely, if $g$ is securely computable by $\cM$, then $H(G)
\leq C(\cM)$.
\end{theorem}
Theorem \ref{t_SC} is, in fact, a special case of our main result
in Theorem \ref{t_GSC} below.
\begin{example}
Let $m = 2$, and let $X_1$ and $X_2$ be $\{0,1\}$-valued rvs with
\begin{align}\nonumber
P_{X_1}(1) &= p = 1- P_{X_1}(0), \quad 0<p<1,\\\nonumber
P_{X_2\mid X_1}(1 \mid 1) &=  P_{X_2\mid X_1}(0 \mid 0) = 1 -
\delta, \quad 0 < \delta < \frac{1}{2}.
\end{align}
Let $g(x_1, x_2) = x_1 + x_2 \mod 2$.

From \cite{Mau93}, \cite{AhlCsi93} (and also Theorem \ref{t_CSK}
above), $C(\{1, 2\}) = h(p \ast \delta) -h(\delta)$, where $p \ast
\delta = (1 - p)\delta + p(1 - \delta)$. Since $H(G) = h(\delta)$,
by Theorem \ref{t_SC} $g$ is securely computable if
\begin{align}\label{e_Ex1i}
2h(\delta) < h(p \ast \delta).
\end{align}
We give a simple scheme for the secure computation of $g$ when $p
= \frac{1}{2}$, that relies on Wyner's well-known method for
Slepian-Wolf data compression \cite{Wyn74} and a derived SK
generation scheme in \cite{Ye05}, \cite{YeTh}. We can write
\begin{align}\label{e_EX1ii}
X_1^n = X_2^n + G^n \,\, \mod 2
\end{align}
with $G^n$ being independent separately of $X^n_2$ and $X^n_1$. We
observe as in \cite{Wyn74} that there exists a binary linear code,
of rate $\cong 1 - h(\delta)$, with parity check matrix
$\mathbf{P}$ such that $X_1^n$, and so $G^n$, is
$\ep_n$-recoverable from $(F_1, X^n_2)$ at terminal 2, where the
Slepian-Wolf codeword $F_1 = \mathbf{P}X^n_1$ constitutes public
communication from terminal 1, and where $\ep_n$ decays to $0$
exponentially rapidly in $n$. Let $\widehat{G^n}$ be the estimate
of $G^n$ thereby formed at terminal 2. Further, let $K = K(X_1^n)$
be the location of $X_1^n$ in the coset of the standard array
corresponding to $\mathbf{P}$. By the previous observation, $K$
too is $\ep_n$-recoverable from $(F_1, X_2^n)$ at terminal 2. From
\cite{Ye05}, \cite{YeTh}, $K$ constitutes a ``perfect" SK for
terminals 1 and 2, of rate $\cong I(X_1 \wedge X_2) = 1-
h(\delta)$, and satisfying
\begin{align}\label{e_Ex1iii}
I(K \wedge F_1) = 0.
\end{align}
Also, observe from (\ref{e_EX1ii}) that $K = K(X^n_1) = K(X^n_2 +
G^n)$ and $F_1 = F_1(X^n_1) = F_1(X^n_2 + G^n)$, and for each
fixed value of $G^n$, the (common) arguments of $K$ and $F_1$ have
the same distribution as $X^n_1$. Hence by (\ref{e_Ex1iii}),
\begin{align}\label{e_Ex1iv}
I(K \wedge F_1, G^n) = I(K \wedge F_1\mid G^n) = 0,
\end{align}
since $I(K \wedge G^n) \leq I(X_1^n \wedge G^n) = 0$.

Then terminal 2 communicates $\widehat{G^n}$ in encrypted form as
$$F_2 = \widehat{G^n} + K \, \mod 2$$ (all represented in bits),
with encryption feasible since $$H(G) = h(\delta) < 1- h(\delta)
\cong \frac{1}{n}H(K),$$ by the sufficient condition
(\ref{e_Ex1i}). Terminal 1 then decrypts $F_2$ using $K$ to
recover $\widehat{G^n}$. The computation of $g^n$ is secure since
\begin{align}\nonumber
I(G^n\wedge F_1, F_2) = I(G^n\wedge F_1) + I(G^n\wedge F_2\mid
F_1)
\end{align}
is small; specifically, the first term equals $0$ since $I(G^n
\wedge F_1) \leq I(G^n  \wedge  X_1^n) = 0$, while the second term
is bounded using (\ref{e_Ex1iv}) according to
\begin{align}\nonumber
I(G^n\wedge F_2\mid F_1) &= H(\widehat{G^n} + K\mid F_1) -
H(\widehat{G^n} + K \mid F_1, G^n)\\\nonumber &\leq H(K) - H(G^n+K
\mid F_1, G^n) + \delta_n\\\nonumber &= I(K \wedge F_1, G^n) +
\delta_n = \delta_n,
\end{align}
where the inequality follows by Fano's inequality and the
exponential decay of $\ep_n$ to $0$. \qed
\end{example}

Next, we turn to the general model for the secure computability of
$g$ by a given set $\cA \subseteq \cM$. Again in the manner of
(\ref{e_SCn}), it is clear that a necessary condition is
\begin{align}\nonumber
H(G) \leq C(\cA).
\end{align}
In contrast, when $\cA \varsubsetneq \cM$, $H(G) < C(\cA)$ is
\emph{not} sufficient for $g$ to be securely computable by $\cA$
as seen by the following simple example.
\begin{example}
Let $m = 3$, $A = \{1,2\}$ and consider rvs $X_1,X_2,X_3$ with
$X_1 = X_2$, where $X_1$ is independent of $X_3$ and $H(X_3) <
H(X_1)$. Let $g$ be defined by $g(x_1,x_2,x_3) = x_3$, $x_i \in
\cX_i$, $1 \leq i \leq 3$. Clearly, $C(\{1,2\}) = H(X_1)$.
Therefore, $H(G) = H(X_3) < C(\{1,2\})$. However, for $g$ to be
computed by the terminals $1$ and $2$, its value must be conveyed
to them necessarily by public communication from terminal $3$.
Thus, $g$ is not securely computable. \qed
\end{example}

Interestingly, the secure computability of $g$ can be examined in
terms of a new SK generation problem that is formulated next.

\subsection{Secret Key Aided by Side Information}
We consider an extension of the SK generation problem in
Definition \ref{d_SK}, which involves additional side information
$Z^n_{\cA'}$ that is correlated with $X^n_\cM$ and is provided to
the terminals in $\cA'$ for use in \emph{only the recovery stage}
of SK generation; however, the public communication $\mathbf{F}$
remains as in Definition \ref{d_PubComm}. Formally, the extension
is described in terms of generic rvs $(X_1,\ldots, X_m,\{Z_i, i
\in \cA'\})$, where the rvs $Z_i$ too take values in finite sets
$\cZ_i$, $i$ in $\cA'$. We note that the full force of this
extension will not be needed to characterize the secure
computability of $g$; an appropriate particularization will
suffice. Nevertheless, this concept is of independent interest.
\begin{definition}
A function $K$ of $(U_\cM, X_\cM^n, Z_{\cA'}^n)$ is an $\ep_n$-
secret key aided by side information $Z_{\cA'}^n$ ($\ep_n$-ASK)
for the terminals $\cA'\subseteq \cM$, $|\cA'|\geq 2$, achievable
from observations of length $n$, randomization $U_\cM$ and public
communication $\mathbf{F}=\mathbf{F}(U_\cM, X_\cM^n)$ if it
satisfies the conditions in Definition \ref{d_SK} with $(U_i,
X_i^n, Z_i^n, \mathbf{F})$ in the role of $(U_i, X_i^n,
\mathbf{F})$ in condition (i). The corresponding ASK capacity
$C(\cA',Z_{\cA'})$ is defined analogously as in Definition
\ref{d_SK}.
\end{definition}
In contrast with the omniscience rate of $H(X_\cM)$ that appears
in the passage following Theorem \ref{t_CSK}, now an underlying
analogous notion of omniscience will involve total common
randomness of rate exceeding $H(X_\cM)$. Specifically, the
enhanced common randomness rate will equal the entropy of the
``maximum common function" (mcf) of the rvs $(X_\cM, Z_i)_{i \in
\cA}$, introduced for a pair of rvs in \cite{GacKor73} (see also
\cite[Problem 3.4.27]{CsiKor81}).

\begin{definition}\label{d_mcf}\cite{GacKor73} For two rvs $Q, R$ with values in
finite sets $\cQ, \cR$, the equivalence relation $q \sim q'$ in
$\cQ$ holds if there exist $N \geq 1$ and sequences $(q_0,
q_1,\ldots, q_N)$ in $\cQ$ with $q_0 = q$, $q_N = q'$ and $(r_1,
\ldots, r_N)$ in $\cR$ satisfying $\bPr{Q = q_{l-1}, R = r_l} > 0$
and $\bPr{Q = q_l, R = r_l} > 0$, $l =1, \ldots, N$. Denote the
corresponding equivalence classes in $\cQ$ by $\cQ_1, \ldots,
\cQ_k$. Similarly, let $\cR_1, \ldots, \cR_{k'}$ denote the
equivalence classes in $\cR$. As argued in \cite{GacKor73}, $k =
k'$ and for $1\leq i,j\leq k$,
\begin{align}\nonumber
\bPr{Q\in \cQ_i\mid R\in \cR_j} = \bPr{R\in \cR_j\mid Q\in \cQ_i}
= \bigg\{ \begin{matrix}1, \quad i=j,\\ 0,\quad i\neq
j.\end{matrix}
\end{align}
The mcf of the rvs $Q, R$ is a rv $\mcf(Q, R)$ with values in
$\{1, \ldots, k\}$ and pmf
\begin{align}\nonumber
\bPr{\mcf(Q, R) =i } = \bPr{Q \in \cQ_i} = \bPr{Q \in \cQ_i, R \in
\cR_i}, \quad i = 1, \ldots, k.
\end{align}
For rvs $Q_1, ..., Q_m$ taking values in finite alphabets, we
define the $\mcf(Q_1, ..., Q_m)$ recursively by
\begin{align}\label{e_mcf:rec}
\mcf(Q_1, ..., Q_m) = \mcf\big(\mcf(Q_1, ..., Q_{m-1}), Q_m\big)
\end{align}
with $\mcf(Q_1, Q_2)$ as above.
\end{definition}
\begin{definition}\label{d_mcfn}
With $Q^n$ denoting $n$ i.i.d. repetitions of the rv $Q$, we
define
\begin{align}\label{e_mcfn}
\mcf^n(Q_1, ...,Q_m)= \left\{\mcf\left(Q_{1t}, ...,
Q_{mt}\right)\right\}_{t = 1}^n.
\end{align}
Note that $\mcf^n(Q_1, ...,Q_m)$ is a function of \emph{each}
individual $Q_i^n, i = 1, ...,m$.
\end{definition}

\begin{remark*} As justification for the definition
(\ref{e_mcf:rec}), consider a rv $\xi$ that satisfies
\begin{align}\label{e_mcf:justification}
H(\xi \mid Q_i) = 0, \quad i = 1,...,m
\end{align}
and suppose for any other rv $\xi'$ satisfying
(\ref{e_mcf:justification}) that $H(\xi) \geq H(\xi')$. Then Lemma
\ref{l_mcf} below shows that $\xi$ must satisfy $H(\xi) =
H(\mcf(Q_1,..., Q_m))$.

\end{remark*}
The following result for the mcf of $m\geq 2$ rvs is a simple
extension of the classic result for $m = 2$ \cite[Theorem
1]{GacKor73}.
\begin{lemma}\label{l_mcf}
Given $0 < \ep <1$, if $\xi^{(n)}$ is $\ep$-recoverable from
$Q_i^n$ for each $i = 1, ..., m$, then
\begin{align}\label{e_mcf:rate bound}
\displaystyle\lim\sup_n \frac{1}{n}H\left(\xi^{(n)}\right) \leq
H(\mcf(Q_1, ..., Q_m)).
\end{align}
\end{lemma}
\noindent\textbf{Proof:} The proof involves a recursive
application of  \cite[Lemma, Section 4]{GacKor73} to $\mcf(Q_1,
..., Q_m)$ in (\ref{e_mcf:rec}), and is provided in  Appendix A.
%
%
%

We are now in a position to characterize ASK capacity. In a manner
analogous to Theorem \ref{t_CSK}, this is done in terms of
$H(\mcf(X_\cM, Z_i)_{i \in \cA'})$ and the smallest rate of
communication $R_{CO}(\cA', Z_{\cA'})$ for each terminal in $\cA'$
to attain omniscience
that corresponds to $n$ i.i.d. repetitions of $\mcf(X_\cM,
Z_i)_{i\in \cA'}$.
\begin{theorem}\label{t_CASK}
The ASK capacity $C(\cA'; Z_{\cA'})$ equals
\begin{align}\nonumber
C(\cA'; Z_{\cA'}) &= H(mcf((X_\cM, Z_i)_{i\in \cA'})) -
R_{CO}(\cA'; Z_{\cA'})
\\\nonumber \text{where} \hspace{2cm}&
\\\nonumber
R_{CO}(\cA';Z_{\cA'}) &= \displaystyle \min_{R_\cM \in \cR(\cA';
Z_{\cA'})}\sum_{i\in \cM} R_i
\\\nonumber \text{with}\hspace{2.25cm}&
\\\label{e_CASK}
\cR(\cA'; Z_{\cA'}) &= \bigg\{ R_\cM :  R_B \geq \displaystyle
\max_{j \in B^c\cap \cA'}H(X_B \mid X_{B^c}, Z_j),\quad B
\varsubsetneq \cM, \cA' \nsubseteq B\bigg\}.
\end{align}
\end{theorem}
The proof of Theorem \ref{t_CASK} is along the same lines as that
of Theorem \ref{t_CSK} \cite{CsiNar04} and is provided in Appendix
B.

The remark following Theorem \ref{t_CSK} also applies to the ASK
capacity $C(\cA';Z_{\cA'})$, as will be seen from the proof of
Theorem \ref{t_CASK}.

\subsection{Characterization of Secure Computability}
If $g$ is securely computable by the terminals in $\cA$, then
$G^n$ constitutes an ASK for $\cM$ under the constraint
(\ref{e_sin''}), of rate $H(G)$, with side information in the form
of $G^n$ provided only to the terminals in $A^c$ in the recovery
stage of SK generation. Thus, a necessary condition for $g$ to be
securely computable by $\cA$, in the manner of (\ref{e_SCn}), is
\begin{align}\label{e_GSCn1}
H(G) &\leq C(\cM; Z_\cM),
\end{align}
where $Z_\cM = Z_\cM(\cA) = \{Z_i\}_{i \in \cM}$ with
\begin{align}
\label{e_ZM} Z_i & = \Bigg\{\begin{matrix} 0, \quad i \in \cA\\G,
\quad i \in \cA^c .\end{matrix}
\end{align}
By particularizing Theorem \ref{t_CASK} to the choice of $Z_\cM$
as above, the right side of  (\ref{e_GSCn1}) reduces to
\begin{align}\label{e_GSCn2}
C(\cM; Z_\cM) &= H(X_\cM) - R_{CO}(\cM; Z_{\cM})
\\\nonumber \text{where}\hspace{2cm}&
\\\nonumber
R_{CO}(\cM;Z_{\cM}) &= \displaystyle \min_{R_\cM \in \cR(\cM;
Z_{\cM})}\sum_{i\in \cM} R_i
\\\nonumber \text{with}\hspace{2.25cm}&
\\\nonumber
\cR(\cM; Z_{\cM}) &= \left\{ R_\cM :  R_B \geq \Bigg\{
\begin{matrix}
H(X_B \mid X_{B^c}),\quad B \varsubsetneq \cM, \cA \nsubseteq B\\
H(X_B \mid X_{B^c}, G),\quad B \varsubsetneq \cM, \cA \subseteq
B\end{matrix}\quad\right\}.
\end{align}
Our main result says that the necessary condition (\ref{e_GSCn1})
is tight.
\begin{theorem}\label{t_GSC}
A function $g$ is securely computable by $\cA \subseteq \cM$ if
\begin{align}\label{e_GSCs}
H(G) < C(\cM; Z_\cM).
\end{align}
Furthermore, under the condition above, $g$ is securely computable
with noninteractive communication and without recourse to
randomization at the terminals in $\cM$.

Conversely, if $g$ is securely computable by $\cA \subseteq \cM$,
then $H(G) \leq C(\cM; Z_\cM)$.
\end{theorem}
\begin{remarks*}
(i) It is easy to see that $C(\cM) \leq C\left(\cM; Z_\cM\right) =
C\left(\cM; Z_\cM(\cA)\right) \leq C(\cA)$. In particular, the
second inequality holds since in the context of $C\left(\cM;
Z_\cM\right)$ the side information for recovery $Z_\cM$ in
(\ref{e_ZM}) is not provided to the terminals in $\cA$ and by
noting that a SK for $\cM$ is also a SK for $\cA$.

\noindent(ii) Observe in Example 2 that $C\left(\cM; Z_\cM\right)
=  C(\cM) = 0$ and so, by Theorem \ref{t_GSC}, $g$ is not securely
computable as noted earlier.
\end{remarks*}
\begin{example}
For the auction example in Section \ref{s_int}, $\cA = \{1, ...,
m-1\}$ and $X_1, ..., X_{m-1}$ are i.i.d. rvs distributed
uniformly on $\{1, ..., k\}$, while $X_m = (X_1, ..., X_{m-1})$.
Let $g_1(x_1, ..., x_m) =\displaystyle \max_{1 \leq i \leq
m-1}x_i$ and $g_2(x_1, ..., x_m) = \displaystyle\arg\max_{1 \leq i
\leq m-1}x_i$. Then, straightforward computation yields for $k<
m-1$ that $$H (G_1) < \log k < H(G_2) =\log (m-1),$$ and for both
$g_1, g_2$ that
$$C\left(\cM; Z_\cM\right) = C(\cM),$$ where, by Theorem \ref{t_CSK}, $$C(\cM) = H(X_\cM) - R_{CO}(\cM) = (m - 1)\log k - (m - 2)\log k = \log k.$$
By Theorem \ref{t_GSC}, $g_1$ is securely computable whereas $g_2$
is not. In fact, $g_2$ is not securely computable by \emph{any}
terminal $i \in \{1, ..., m-1\}$. This, too, is implied by Theorem
\ref{t_GSC} upon nothing that for each $i \in \{1, ..., m-1\}$ and
a restricted choice $\cA = \{i\}$, $$C\left(\cM; Z_\cM(\cA)\right)
= H(X_i) = \log k < \log (m-1) = H(G_2),$$ where the first
equality is a consequence of remark (i) following Theorem
\ref{t_GSC} and remark (ii) after Definition \ref{d_SK}. \qed
\end{example}

\subsection{A Decomposition Result}
The sufficiency condition (\ref{e_GSCs}) prompts the following two
natural questions: Does the difference $C\left(\cM; Z_\cM \right)
-  H(G)$ possess an operational significance? If $g$ is securely
computable by the terminals in $\cA$, clearly $G^n$ forms a SK for
$\cA$. Can $G^n$ be augmented suitably to form a $SK$ for $\cA$ of
maximum achievable rate?

The answers to both these questions are in the affirmative. In
particular, our approach to the second question involves a
characterization of the minimum rate of communication for
omniscience for $\cA$, under the additional requirement that this
communication be independent of $G^n$. Specifically, we show below
that for a securely computable function $g$, this minimum rate
remains $R_{CO}(\cA)$ (see (\ref{e_RCO})).

Addressing the first question, we introduce a rv $K_g = K^{(n)}_g$
such that $K = \left(K_g, G^n\right)$ constitutes an $\ep_n$-ASK
for $\cM$ with side information $Z_\cM$ as in (\ref{e_ZM}) and
satisfying the additional requirement
\begin{align}\label{e_Kg}
I\left(K_g \wedge G^n\right) \leq \ep_n.
\end{align}
Let the largest rate $\lim_n (1/n)  \log |\cK_g^{(n)}| $ of such
an ASK be $C^g\left(\cM; Z_\cM\right)$. Observe that since $K$ is
required to be nearly independent of $\mathbf{F}$, where
$\mathbf{F}$ is the public communication involved in its
formation, it follows by (\ref{e_Kg}) that $K_g$ is nearly
independent of $\left(G^n, \mathbf{F}\right)$.

Turning to the second question, in the same vein let $K_g'$ be a
rv such that $K' = \left(K_g', G^n\right)$ constitutes an
$\ep_n$-SK for $\cA \subseteq \cM$ and satisfying (\ref{e_Kg}).
Let $C^g(\cA)$ denote the largest rate of $K_g'$. As noted above,
$K_g'$ will be nearly independent of $(G^n, \mathbf{F}')$, where
$\mathbf{F}'$ is the public communication involved in the
formation of $K'$.

\begin{proposition}\label{p_CSg}
For $\cA \subseteq \cM$, it holds that
\begin{align}\nonumber
(i)\quad C^g\left(\cM; Z_\cM(\cA)\right) &= C\left(\cM;
Z_\cM(\cA)\right) - H(G),\\\nonumber (ii)\quad\qquad \qquad
C^g(\cA) &= C(\cA) - H(G).
\end{align}
\end{proposition}
\begin{remarks*}
\noindent(i) For the case $\cA = \cM$, both (i) and (ii) above
reduce to $C^g(\cM) = C(\cM) - H(G)$.

\noindent(ii) Theorem \ref{t_CSK} and Proposition
\ref{p_CSg}\,(ii) lead to the observation
\begin{align}\nonumber
H(X_\cM) = R_{CO}(\cA) + H(G) + C^g(\cA),
\end{align}
which admits the following heuristic interpretation. The  ``total
randomness" $X_\cM^n$ that corresponds to omniscience decomposes
into three  ``nearly mutually independent" components: a
minimum-sized communication for omniscience for $\cA$ and the
independent parts of an optimum-rate SK for $\cA$ composed of
$G^n$ and $K_g'$.


\end{remarks*}

\section{Proofs of Theorem \ref{t_GSC} and Proposition \ref{p_CSg}}\label{s_pro}

\subsection{Proof of Theorem \ref{t_GSC}}

The necessity of (\ref{e_GSCn1}) follows by the comments preceding
Theorem \ref{t_GSC}.

The sufficiency of (\ref{e_GSCs}) will be established by showing
the existence of \emph{noninteractive} public communication
comprising source codes that enable omniscience corresponding to
$X^n_\cM$ at the terminals in $\cA$, and thereby the computation
of $g$. Furthermore, the corresponding codewords are selected so
as to be simultaneously independent of $G^n$, thus assuring
security.

First, from (\ref{e_GSCs}) and (\ref{e_GSCn2}), there exists
$\delta > 0$ such that $ R_{CO}(\cM; Z_\cM) + \delta < H(X_\cM|
G)$, using $G = g(X_\cM)$. For each $i$ and $R_i \geq 0$, consider
a (map-valued) rv $J_i$ that is uniformly distributed on the
family $\cJ_i$ of all mappings $\cX^n_i \rightarrow \{1, \ldots,
\lceil\exp(nR_i)\rceil\},$ $i \in \cM$. The rvs $J_1, ..., J_m,
X^n_\cM$ are taken to be mutually independent.

Fix $\ep, \ep'$, with $\ep' > m\ep$ and $\epsilon + \epsilon' <
1$. It follows from the proof of the general source network coding
theorem \cite[Lemma 3.1.13 and Theorem 3.1.14]{CsiKor81} that for
all sufficiently large $n$,
\begin{align}\nonumber
\bPr{\left\{j_\cM \in \cJ_\cM: X^n_\cM \text{ is }
\ep_n\text{-recoverable from }
\left(X^n_i,j_{\cM\backslash\{i\}}\left(X^n_{\cM\backslash\{i\}}\right),
Z_i^n\right), i \in \cM\right\}}\\\label{e_SC:rel}
&\hspace{-1cm}\geq 1 - \ep,
\end{align}
provided $R_\cM = (R_1, ..., R_m) \in \cR(\cM; Z_\cM)$, where
$\ep_n$ vanishes exponentially rapidly in $n$. This assertion
follows exactly as in the proof of \cite[Proposition 1, with $A =
\cM$]{CsiNar04} but with $\tilde{X}_i$ there equal to $(X_i, Z_i)$
rather than $X_i$, $i \in \cM$. In particular, we shall choose
$R_\cM \in \cR(\cM;Z_\cM)$ such that
\begin{align}\label{e_SC:R_M}
\sum_{i=1}^m R_i \leq  R_{CO}(\cM; Z_\cM) + \frac{\delta}{2}.
\end{align}

Below we shall establish that
\begin{align}\label{e_SC:sec}
\bPr{\left\{j_\cM \in \cJ_\cM: I\left(j_\cM(X^n_\cM) \wedge
G^n\right) \geq \ep_n\right\}} \leq \ep',
\end{align}
for all $n$ sufficiently large, to which end it suffices to show
that
\begin{align}\label{e_SC:bsec}
\bPr{\left\{j_\cM \in \cJ_\cM: I\left(j_i(X^n_i) \wedge G^n,
j_{\cM\backslash\{i\}}\left(X^n_{\cM\backslash\{i\}}\right)\right)
\geq \frac{\ep_n}{m}\right\}} \leq \frac{\ep'}{m}, \quad i \in
\cM,
\end{align}
since
\begin{align}\nonumber
I\left(j_\cM\left(X^n_\cM\right) \wedge G^n\right) &= \sum_{i=1}^m
I\left(j_i\left(X^n_i\right)\wedge G^n \mid j_1\left(X^n_1\right),
\ldots, j_{i-1}\left(X^n_{i-1}\right)\right)\\\nonumber &\leq
\sum_{i=1}^m I\left(j_i\left(X^n_i\right)\wedge G^n,
j_{\cM\backslash\{i\}}\left(X^n_{\cM\backslash\{i\}}\right)\right).
\end{align}
Then it would follow from (\ref{e_SC:rel}), (\ref{e_SC:sec}) and
definition of $Z_\cM$ in (\ref{e_GSCn1}) that
\begin{align}\nonumber
\mathtt{Pr}\bigg(\bigg\{ j_\cM \in \cJ_\cM: G^n \text{ is }
\ep_n\text{-recoverable from }&\left(X^n_i,
j_{\cM\backslash\{i\}}\left(X^n_{\cM\backslash\{i\}}\right)\right),
i \in \cA,\\\nonumber \text{and }&I(j_\cM(X^n_\cM) \wedge G^n) <
\ep_n\bigg\}\bigg)\geq 1 - \ep - \ep'.
\end{align}
This shows the existence of a particular realization $j_\cM$ of
$J_\cM$ such that $G^n$ is $\ep_n$-SC from \\$(X^n_i,
j_{\cM\backslash \{i\}}\left(X^n_{\cM\backslash \{i\}}\right))$
for each $i \in \cA$.

It now remains to prove (\ref{e_SC:bsec}). Fix $i \in \cM$ and
note that for each $j_i \in \cJ_i$, with $\|j_i\|$ denoting the
cardinality of the (image) set $j_i(\cX_i^n)$,
\begin{align}\nonumber
I\big(j_i&\left(X^n_i\big) \wedge G^n,
j_{\cM\backslash\{i\}}\left(X^n_{\cM\backslash\{i\}}\right)\right)
\\\nonumber&\leq I\left(j_i(X^n_i) \wedge G^n,
j_{\cM\backslash\{i\}}\left(X^n_{\cM\backslash\{i\}}\right)\right)
+ \log \|j_i\| - H\left(j_i(X^n_i)\right)
\\\label{e_SC:sin} &= D(j_i(X^n_i),(G^n, j_{\cM\backslash\{i\}}(X^n_{\cM\backslash\{i\}}) \|
U_{j_i(\cX_i^n)}\times \left(G^n,
j_{\cM\backslash\{i\}}\left(X^n_{\cM\backslash\{i\}}\right)\right),
\end{align}
where the right side above denotes the (Kullback-Leibler)
divergence between the joint pmf of\\
$j_i(X^n_i)$,$\left(G^n,j_{\cM\backslash\{i\}}\left(X^n_{\cM\backslash\{i\}}\right)\right)$
and the product of the uniform pmf on $j_i(\cX_i^n)$ and the pmf
of \\$\left(G^n,
j_{\cM\backslash\{i\}}\left(X^n_{\cM\backslash\{i\}}\right)\right)$.
Using \cite[Lemma 1]{CsiNar04}, the right side of (\ref{e_SC:sin})
is bounded above further by
\begin{align}
s_{var}\log \frac{\|j_i\|}{s_{var}},
\end{align}
where $s_{var} = s_{var}(j_i(X^n_i); G^n,
j_{\cM\backslash\{i\}}(X^n_{\cM\backslash\{i\}})$ is the
variational distance between the pmfs in the divergence above.
Therefore, to prove (\ref{e_SC:bsec}), it suffices to show that
\begin{align}\label{e_SC:bbsec}
\bPr{\left\{j_\cM \in \cJ_\cM: s_{var}\left(j_i(X^n_i); G^n,
j_{\cM\backslash\{i\}}\left(X^n_{\cM\backslash\{i\}}\right)\right)
\geq \frac{\ep_n}{m}\right\}} \leq \frac{\ep'}{m},\quad i \in \cM,
\end{align}
on account of the fact that $\log\|j_i(X^n_i)\| = O(n)$, and the
exponential decay to $0$ of $\ep_n$. Defining
\begin{align}\nonumber
\tilde{\cJ}_i = \left\{j_{\cM\backslash\{i\}} \in
\cJ_{\cM\backslash\{i\}}: X_\cM^n\text{ is
}\ep_n\text{-recoverable from
}\left(X^n_i,j_{\cM\backslash\{i\}}\left(X^n_{\cM\backslash\{i\}}\right),
Z^n_i\right)\right\},
\end{align}
we have by (\ref{e_SC:rel}) that $\bPr{J_{\cM\backslash\{i\}} \in
\tilde{\cJ}_i} \geq 1- \ep$. Thus, in (\ref{e_SC:bbsec}),
\begin{align}\nonumber
&\bPr{\left\{j_\cM \in \cJ_\cM:
s_{var}\left(j_i\left(X^n_i\right); G^n,
j_{\cM\backslash\{i\}}\left(X^n_{\cM\backslash\{i\}}\right)\right)
\geq \frac{\ep_n}{m}\right\}}\\\nonumber &\leq \ep +
\sum_{j_{\cM\backslash\{i\}} \in
\tilde{\cJ}_i}\bPr{J_{\cM\backslash\{i\}} = j_{\cM\backslash\{i\}}
}\times\\\nonumber &\hspace{2.5cm}\bPr{\left\{j_i \in \cJ_i:
s_{var}\left(j_i(X^n_i); G^n,
j_{\cM\backslash\{i\}}\left(X^n_{\cM\backslash\{i\}}\right)\right)\geq
\frac{\ep_n}{m}\right\}},
\end{align}
since $J_i$ is independent of $J_{\cM\backslash\{i\}}$. Thus,
(\ref{e_SC:bbsec}), and hence (\ref{e_SC:bsec}), will follow upon
showing that
\begin{align}\label{e_SC:bbbsec}
\bPr{\left\{j_i \in \cJ_i: s_{var}\left(j_i(X^n_i); G^n,
j_{\cM\backslash\{i\}}\left(X^n_{\cM\backslash\{i\}}\right)\right)\geq
\frac{\ep_n}{m}\right\}} \leq \frac{\ep'}{m} - \ep,\quad
j_{\cM\backslash\{i\}} \in \tilde{\cJ}_i,
\end{align}
for all $n$ sufficiently large. Fix $j_{\cM\backslash\{i\}} \in
\tilde{\cJ}_i$. We take recourse to Lemma \ref{l_B} in Appendix C,
and set $U = X^n_\cM, U' = X^n_i, V = G^n, h =
j_{\cM\backslash\{i\}}$, and
\begin{align}\nonumber
\cU_0 = \left\{x_\cM^n \in \cX^n_\cM : x^n_\cM =
\psi_i\left(x^n_i,
j_{\cM\backslash\{i\}}\left(x^n_{\cM\backslash\{i\}}\right),
g^n\left(x_\cM^n\right)\mathbf{1}\left(i \in \cA^c\right)\right)
\right\}
\end{align}
for some mapping $\psi_i$. By the definition of $\tilde{\cJ}_i$,
\begin{align}\nonumber
\bPr{U \in \cU_0} \geq 1 - \ep_n,
\end{align}
so that condition (\ref{e_bound0})(i) preceding Lemma \ref{l_B} is
met. Condition (\ref{e_bound0})(ii), too, is met since conditioned
on the events in (\ref{e_bound0})(ii), only those $x_\cM^n \in
\cU_0$ can occur that are determined uniquely by their $i^{th}$
components $x_i^n$.

Upon choosing
\begin{align}\nonumber
d = \exp\left[n\left(H(X_\cM | G) -
\frac{\delta}{6}\right)\right],
\end{align}
in (\ref{e_bound1}), the hypotheses of Lemma \ref{l_B} are
satisfied with $\lambda = \sqrt{\ep_n}$, for an appropriate
exponentially vanishing $\ep_n$. Then, by Lemma \ref{l_B}, with
\begin{align}\nonumber
r = \left\lceil\exp[nR_i]\right\rceil,\quad r' =
\left\lceil\exp\left[n\left(\sum_{l \in \cM\backslash\{i\}} R_l +
\frac{\delta}{6}\right)\right]\right\rceil,
\end{align}
and with $J_i$ in the role of $\phi$, we get from (\ref{e_bc}) and
(\ref{e_SC:R_M}) that
\begin{align}\nonumber
\bPr{\left\{j_i \in \cJ_i: s_{var}\left(j_i(X^n_i); G^n,
j_{\cM\backslash\{i\}}\left(X^n_{\cM\backslash\{i\}}\right)\right)\geq
14\sqrt{\ep_n}\right\}}
\end{align}
decays to $0$ doubly exponentially in $n$, which proves
(\ref{e_SC:bbbsec}). This completes the proof of Theorem
\ref{t_GSC}. \qed

\subsection{Proof of Proposition \ref{p_CSg}} (i) Since the rv
$(K_g^{(n)}, G^n)$, with nearly independent components,
constitutes an ASK for $\cM$ with side information $Z_\cM$ as in
(\ref{e_ZM}), it is clear that \begin{align}\label{e_propInq1}
H(G) + C^g\left(\cM; Z_\cM\right) \leq C\left(\cM; Z_\cM\right).
\end{align} In order to prove the reverse of (\ref{e_propInq1}),
we show that $C\left(\cM; Z_\cM\right) - H(G)$ is an achievable
ASK rate for $K_g$ that additionally satisfies (\ref{e_Kg}).
First, note that in the proof of Theorem \ref{t_GSC}, the
assertions (\ref{e_SC:rel}) and (\ref{e_SC:bsec}) mean that for
all sufficiently large $n$, there exists a public communication
$F_\cM$, say, such that $I(F_\cM \wedge G^n) < \ep_n$ and
$X^n_\cM$ is $\ep_n$-recoverable from $(X_i^n, F_\cM, Z_i^n)$ for
every $i \in \cM$, with $\displaystyle \lim_n \ep_n =0$. Fix $0 <
\tau < \delta$, where $\delta$ is as in the proof of Theorem
\ref{t_GSC}. Apply Lemma \ref{l_B}, choosing
\begin{align}\label{e_applybc2Kg}
U = U' = X^n_\cM, \quad \cU_0 = \cX_\cM^n, \quad V = G^n,\quad h =
F_\cM, \quad d = \exp\left[n \left(H\left(X_\cM | G\right)-
\frac{\tau}{6}\right)\right],
\end{align}
whereby the hypothesis (\ref{e_bound1}) of Lemma \ref{l_B} is
satisfied for all $n$ sufficiently large. Fixing
\begin{align}\nonumber
r' = \left\lceil\exp \left[n \left(R_{CO}\left(\cM; Z_\cM \right)
+ \frac{\tau}{2}\right)\right]\right\rceil,
\end{align}
by Lemma \ref{l_B} a randomly chosen $\phi$ of rate
$$ \frac{1}{n}\log{r} = H(X_\cM | G) - R_{CO} \left(\cM; Z_\cM\right)-
\tau = C\left(\cM; Z_\cM\right) - H(G) - \tau$$ will yield an ASK
$K_g = K_g^{(n)}= \phi\left(X_\cM^n\right)$ which is nearly
independent of $(F_\cM, G^n)$ (and, in particular, satisfies
(\ref{e_Kg})) with positive probability, for all $n$ sufficiently
large.

\noindent(ii) The proof can be completed as that of part (i) upon
showing that for a securely computable $g$, for all $\tau > 0$ and
$n$ sufficiently large, there exists a public communication
$F_\cM'$ that meets the following requirements: its rate does not
exceed $R_{CO}(\cA) + \tau$; $I(F_\cM' \wedge G^n) < \ep_n$; and
$X^n_\cM$ is $\ep_n$-recoverable from $(X_i^n, F_\cM')$ for every
$i \in \cA$. To that end, for $R_\cM = \left(R_1, ..., R_m\right)
\in \cR(\cM ; Z_\cM)$ as in the proof of Theorem \ref{t_GSC},
consider $R_\cM' = \left(R_1', ..., R_m'\right) \in \cR(\cA)$ that
satisfies $R_i' \leq R_i$ for all $i \in \cM$ and
\begin{align}\nonumber
\sum_{i=1}^m R_i' \leq R_{CO}(\cA) + \tau,
\end{align}
noting that $\cR\left(\cM; Z_\cM\right)\subseteq \cR(\cA)$.
Further, for $J_\cM$ and $\cJ_\cM$ as in that proof, define a
(map-valued) rv $J_i'$ that is uniformly distributed on the family
$\cJ_i'$ of all mappings from \\$\{1, \ldots,
\lceil\exp(nR_i)\rceil\}$ to $\{1, \ldots,
\lceil\exp(nR_i')\rceil\}$, $i \in \cM$. The random variables
$J_1, ..., J_m,$\\$ J_1', ..., J_m', X_\cM^n$ are taken to be
mutually independent. Define $\cJ_\cM^0$ as the set of mappings
$j_\cM \in \cJ_\cM$ for which there exists a $j_\cM' \in \cJ_\cM'$
such that $X_\cM^n$ is $\ep_n$-recoverable from \\$\left(X_i^n,
j_\cM'\left(j_\cM \left(X^n_\cM\right)\right)\right)$ for every $i
\in \cA$. By the general source network coding theorem \cite[Lemma
3.1.13 and Theorem 3.1.14]{CsiKor81}, applied to the random
mapping $J_\cM'\left(J_\cM\right)$, it follows that for all
sufficiently large $n$, $$\bPr{ J_\cM \in J_\cM^0 } \geq 1 -
\ep.$$ This, together with (\ref{e_SC:rel}) and (\ref{e_SC:bsec})
in the proof of Theorem \ref{t_GSC}, imply that for a securely
computable $g$ there exist $j_\cM \in \cJ_\cM$ and $j_\cM' \in
\cJ_\cM'$ for which the public communication $F_\cM' \triangleq
j_\cM'(j_\cM)$ satisfies the aforementioned requirements. Finally,
apply Lemma \ref{l_B} with $U, U', \cU_0, V$ and $d$ as in
(\ref{e_applybc2Kg}) but with $h = F_\cM'$ and
$$r' = \left\lceil\exp \left[n \left(R_{CO}\left(\cA \right)
+ \frac{\tau}{2}\right)\right]\right\rceil.$$ As in the proof
above of part (i), a SK $K_g' = K_g'^{(n)}$ of rate
$$ \frac{1}{n}\log{r} = H(X_\cM | G) - R_{CO} \left(\cA\right)-
\tau = C\left(\cA\right) - H(G) - \tau$$ which is nearly
independent of $(F_\cM', G^n)$ (and, hence, satisfies
(\ref{e_Kg})) exists for all $n$ sufficiently large.
 \qed

\section{Discussion}\label{s_dis}
We obtain simple necessary and sufficient conditions for secure
computability involving function entropy and ASK capacity. The
latter is the largest rate of a SK for a new model in which side
information is provided for use in only the recovery stage of SK
generation. This model could be of independent interest. In
particular, a function is securely computable if its entropy is
less than ASK capacity of an associated secrecy model. The
difference is shown to correspond to the maximum achievable rate
of an ASK which is independent of the securely computed function
and, together with it, forms an ASK of optimum rate. Also, a
function that is securely computed by $\cA$ can be augmented to
form a SK for $\cA$ of maximum rate.

Our results extend to functions defined on a block of symbols of
\emph{fixed} length in an obvious manner by considering larger
alphabets composed of supersymbols of such length. However, they
do not cover functions of symbols of increasing length (in $n$).

In our proof of Theorem \ref{t_GSC}, g was securely computed from
omniscience at all the terminals in $\cA \subseteq \cM$ that was
attained using noninteractive public communication. However, as
Example 1 illustrates, omniscience is not necessary for the secure
computation of $g$, and it is possible to make do with
communication of rate less than $R_{CO}(\cM)$ using an interactive
protocol. A related unresolved question is: what is the minimum
rate of public communication for secure computation?

A natural generalization of the conditions for secure
computability of $g$ by $\cA \subseteq \cM$ given here entails a
characterization of conditions for the secure computability of
multiple functions $g_1, ..., g_k$ by $\cA_1,..., \cA_k$ of $\cM$,
respectively. This unsolved problem, in general, will not permit
omniscience for any $\cA_i, i = 1, ..., k$. For instance with $m =
2$, $\cA_1 = \{1\}$, $\cA_2 = \{2\}$, and $X_1$ and $X_2$ being
independent, the functions $g_i(x_i) = x_i$, $i = 1,2$, are
securely computable trivially, but not through omniscience since,
in this example, public communication is forbidden for the secure
computation of $g_1, g_2$.

\section*{Appendix A}
\setcounter{equation}{0}
\renewcommand{\theequation}{A\arabic{equation}}
The proof of Lemma \ref{l_mcf} is based on \cite[Lemma, Section
4]{GacKor73}, which is paraphrased first. Let the rvs $Q$ and $R$
take values in the finite set $\cQ$ and $\cR$, respectively. For a
stochastic matrix $W: \cQ \rightarrow \cQ$, let $\{\tilde{\cD_1},
..., \tilde{\cD_l}\}$ be the ergodic decomposition (into
communicating classes) (cf. e.g., \cite{Lov55}) of $\cQ$ based on
$W$. Let $\tilde{\cD}^{(n)}$ denote a fixed ergodic class of
$\cQ^n$ (the $n$-fold Cartesian product of $\cQ$) on the basis of
$W^n$ (the $n$-fold product of $W$). Let $\cD^{(n)}$ and
$\cR^{(n)}$ be any (nonempty) subsets of $\tilde{\cD}^{(n)}$ and
$\cR^n$, respectively.
\begin{lemmaGK}\cite{GacKor73} For $\tDn, \cDn, \cRn$ as above, assume that
\begin{align}\nonumber
\bPr{Q^n \in \cDn \mid R^n \in \cRn} &\geq \exp[-n
\ep_n],\\\label{e_GK1} \bPr{R^n \in \cRn \mid Q^n \in \cDn} &\geq
\exp[-n \ep_n],
\end{align}
where $\displaystyle \lim_n \ep_n =0$. Then (as stated in
\cite[bottom of p. 157]{GacKor73}),
\begin{align}\label{e_GK2}
\frac{\bPr{Q^n \in \cDn}}{\bPr{Q^n \in \tDn}} \geq \exp[
-n\kappa\ep_n\log^2\ep_n],
\end{align}
for a (positive) constant $\kappa$ that depends only on the pmf of
$(Q, R)$ and on $W$.
\end{lemmaGK}
A simple consequence of (\ref{e_GK2}) is that for a given ergodic
class $\tDn$ and disjoint subsets $\cDn_1, ..., \cDn_t$ of it, and
subsets $\cRn_1, ..., \cRn_t$ (not necessarily distinct) of
$\cR^n$, such that $\cDn_{t'}, \cRn_{t'}, t' = 1, ..., t$, satisfy
(\ref{e_GK1}), then
\begin{align}\label{e_GKbound}
t  \leq \exp[n\kappa \ep_n\log^2\ep_n].
\end{align}
Note that the ergodic decomposition of $Q^n$ on the basis of $W^n$
for the specific choice
\begin{align}\nonumber
W(q|q') = \sum_{r\in \cR}\bPr{Q = q \mid R = r} \bPr{ R = r \mid Q
= q'}, \quad q,q'\in \cQ
\end{align}
corresponds to the set of values of $\mcf^n(Q,R)$ defined by
(\ref{e_mcfn}) \cite{GacKor73}.
Next, pick $Q = Q_m$, $R = (Q_1, ..., Q_{m-1})$, 
and define the stochastic matrix $W : \cQ \rightarrow \cQ$ by
\begin{align}\nonumber
W(q|q') = \sum_\alpha\bPr{Q = q \mid \mcf(Q_1, ..., Q_{m-1}) =
\alpha}& \bPr{\mcf(Q_1, ..., Q_{m-1}) = \alpha \mid Q =
q'},\\\label{e_stochW} &\hspace{3.8cm} q,q'\in \cQ.
\end{align}
The ergodic decomposition of $\cQ^n$ on the basis of $W^n$ (with
$W$ as in (\ref{e_stochW})) will correspond to the set of values
of $\mcf^n(Q_1, ..., Q_m)$, recalling (\ref{e_mcf:rec}). Since
$\xi^{(n)}$ is $\ep$-recoverable from $Q_i^n, i =1, ..., m$, note
that
\begin{align}\nonumber
\xi'^{(n)} = \left(\xi^{(n)}, \mcf^n(Q_1, ..., Q_m)\right)
\end{align}
also is $\ep$-recoverable in the same sense, recalling definition
\ref{d_mcfn}. This implies the existence of mappings $\xi'^{(n)}
_i, i = 1, ..., m$, satisfying
\begin{align}\label{e_recov}
\bPr{\xi'^{(n)}_1(Q_1^n) = ... = \xi'^{(n)}_m(Q_m^n) = \xi'^{(n)}}
\geq 1 - \ep.
\end{align}
For each fixed value $c = (c_1, c_2)$ of $\xi'^{(n)}$, let
\begin{align}\nonumber
\cDn_c &= \left\{ q_m^n \in \cQ^n_m : \xi'^{(n)}_m(q_m^n) = c
\right\},\\\nonumber \cRn_c &= \left\{(q_1^n, ..., q_{m-1}^n) \in
\cQ^n_1\times...\times\cQ^n_{m-1} : \xi'^{(n)}_i(q_i^n) = c, i =
1, ..., m-1 \right\}.
\end{align}
Let $C(\ep)$ denote the set of $c$'s such that
\begin{align}\nonumber
\bPr{Q^n \in \cDn_c \mid R^n \in \cRn_c} &\geq 1 - \sqrt{\ep},
\\\label{e_GK1'} \bPr{R^n \in \cRn_c \mid Q^n \in \cDn_c} &\geq 1
- \sqrt{\ep}.
\end{align}
Then, as in \cite[Proposition 1]{GacKor73}, it follows from
(\ref{e_recov}) that
\begin{align}\label{e_Ce}
\bPr{\xi'^{(n)} \in C(\ep)} &\geq 1 - 4\sqrt{\ep}.
\end{align}
Next, we observe for each fixed $c_2$, that the disjoint sets
$\cDn_{c_1,c_2}$ lie in a fixed ergodic class of $\cQ^n$
(determined by $c_2$). Since (\ref{e_GK1'}) are compatible with
the assumption (\ref{e_GK1})  for all $n$ sufficiently large, we
have from (\ref{e_GKbound}) that
\begin{align}\label{e_boundCe}
\|\{ c_1 : (c_1, c_2) \in C(\ep)\}\| \leq \exp[ n \kappa \ep_n
\log^2 \ep_n ],
\end{align}
where $\kappa$ depends on the pmf of $(Q_1, ...,Q_m)$ and $W$ in
(\ref{e_stochW}), and where $\displaystyle \lim_n \ep_n = 0$.
Finally,
\begin{align}\nonumber
\frac{1}{n} H\left(\xi'^{(n)}\right) &= \frac{1}{n}
H\left(\xi^{(n)}, \mcf^n(Q_1, ..., Q_m\right)
\\\nonumber
&\leq H\left(\mcf(Q_1, ..., Q_m)\right) +
\frac{1}{n}H\left(\xi^{(n)}, \mathbf{1}\left(\xi'^{(n)} \in C(\ep)
\right) \mid \mcf^n(Q_1, ..., Q_m)\right)
\\\nonumber
&= H\left(\mcf(Q_1, ..., Q_m)\right) +  \frac{1}{n}
\\\nonumber &\quad+
\frac{1}{n}H\left(\xi^{(n)} \mid \mcf^n(Q_1, ..., Q_m),
\mathbf{1}\left(\xi'^{(n)} \in C(\ep) \right)\right) \\\nonumber
&\leq H(\mcf(Q_1, ..., Q_m)) + \delta_n,
\end{align}
where $\displaystyle \lim_n \delta_n = 0$ by  (\ref{e_Ce}) and
(\ref{e_boundCe}). \qed

\section*{Appendix B}
\setcounter{equation}{0}
\renewcommand{\theequation}{B\arabic{equation}}
Considering first the achievability part, fix $\delta > 0$. From
the result for a general source network \cite[Theorem
3.1.14]{CsiKor81} it follows, as in the proof of \cite[Proposition
1]{CsiNar04}, that for $R_\cM \in \cR\left(\cA', Z_{\cA'}\right)$
and all $n$ sufficiently large, there exists a noninteractive
communication $\mathbf{F^{(n)}} = (F_1^{(n)}, ..., F_m^{(n)})$
with
\begin{align}\nonumber
\frac{1}{n}\log \|\mathbf{F}^{(n)}\| \leq \sum_{i = 1}^m R_i +
\delta,
\end{align}
such that $\cX^n_\cM$ is $\ep_n$-recoverable from $\left(X^n_i,
Z_i^n, \mathbf{F^{(n)}}\right), i \in \cA'$. Therefore,
$\left\{\mcf \left( (X_{\cM t}, Z_{it})_{i \in
A'}\right)\right\}_{t  = 1}^n$ is $\ep_n$-recoverable from
$\left(X^n_i, Z_i^n, \mathbf{F^{(n)}}\right), i \in \cA'$. The
last step takes recourse to Lemma \ref{l_B} in Appendix C.
Specifically, choose $U = U' = \left\{\mcf \left( (X_{\cM t},
Z_{it})_{i \in A'}\right)\right\}_{t  = 1}^n$, $\cU_0 = \cU$, $V =
\text{constant}$, $h = F^{(n)}$, $d = n\left[H\left(\mcf \left(
(X_\cM, Z_i)_{i \in A'}\right) \right) - \delta \right]$, whereby the hypothesis (\ref{e_bound1}) of Lemma \ref{l_B} is
satisfied for all $n$ sufficiently large. Fixing
\begin{align}\nonumber
r' = \left\lceil\exp \left[n \left(\sum_{i = 1}^m R_i +
\delta\right)\right]\right\rceil,
\end{align}
Lemma \ref{l_B} implies the existence of a $\phi$, and thereby an ASK $K^{(n)} = \phi \left(\left\{\mcf \left( (X_{\cM t},
Z_{it})_{i \in A'}\right)\right\}_{t  = 1}^n\right)$, of rate
$$\frac{1}{n}\log r =  H\left(\mcf \left( (X_\cM, Z_i)_{i \in A'}\right) \right) - \sum_{i = 1}^mR_i  - 3\delta.$$
In particular, we can choose $$\sum_{i = 1}^m R_i \leq
R_{CO}\left(\cA'; Z_{\cA'}\right) + \frac{\delta}{2}.$$ Since
$\delta$ was arbitrary, this establishes the achievability part.

We prove the converse part under either of the weaker conditions
(\ref{e_sin'}) or (\ref{e_sin''}). Let $K = K^{(n)}\left(U_\cM,
X_\cM^n, Z^n_\cM\right)$ be an $\ep_n$-ASK for $\cA'$, achievable
using observations of length $n$, randomization $U_\cM$, public
communication $\mathbf{F} = \mathbf{F}\left(U_\cM, X_\cM^n\right)$
and side information $Z_\cM^n$. Then,
\begin{align}\label{e_ASK1}
\frac{1}{n} H (K ) \leq \frac{1}{n}H(K \mid \mathbf{F}) + \ep_n.
\end{align}
Let $K_u = K\left(u, X_\cM^n, Z^n_\cM\right)$ denote the random
value of the ASK for a fixed $U_\cM = u$. Since $\left(X_\cM^n,
K\right)$ is $\ep_n$-recoverable from the rvs $\left(U_\cM,
X^n_\cM, Z^n_i\right)$ for each $i \in \cA'$,
\begin{align}\nonumber
&\bP{U_\cM}{\left\{u : \left(X_\cM^n, K_u\right) \text{ is
}\sqrt{\ep_n}\text{-recoverable from $\left(U_\cM=u, X^n_\cM,
Z^n_i\right)$ for each $i \in \cA'$ }\right\}}
\\\label{e_U}&\hspace{5in} \geq 1 -\sqrt{\ep_n}.
\end{align}
Also, for each $U_\cM = u$
\begin{align}\nonumber
\frac{1}{n} H\left( X_\cM^n, K \mid U_\cM = u\right) = \frac{1}{n}H\left(X_\cM^n, K_u\right)
\end{align}
by  independence of $U_\cM$ and $\left(X^n_\cM, Z^n_\cM\right)$, and therefore, by Lemma \ref{l_mcf}, for $u$ in the set in (\ref{e_U}),
\begin{align}\label{e_mcfbound}
\frac{1}{n} H\left( X_\cM^n, K \mid U_\cM = u\right) \leq H\left(\mcf\left((X_\cM, Z_i)_{i \in
\cA'}\right)\right) + \delta_n,
\end{align}
for all $n$ sufficiently large and where $\displaystyle \lim_n \delta_n = 0$.
Then,
\begin{align}\label{e_ASK2}
\frac{1}{n} H (U_\cM, X_\cM^n, K ) \leq
\frac{1}{n}H\left(U_\cM\right) + H\left(\mcf\left((X_\cM, Z_i)_{i \in
\cA'}\right)\right) + \delta_n + \sqrt{\ep_n}\log \left(|\cX_\cM||\cZ_\cM|\right),
\end{align}
by (\ref{e_U}) and (\ref{e_mcfbound}).
The proof is now completed along the lines of
\cite[Lemma 2 and Theorem 3]{CsiNar04}. Specifically, denoting the set of
positive integers $\{1, ..., l\}$ by $[1, l]$,
\begin{align}\nonumber
\frac{1}{n} H (U_\cM, X_\cM^n, K ) = \frac{1}{n}H(K \mid \mathbf{F}) +
\sum_{i = 1}^m R_i' + \frac{1}{n}H(U_\cM),
\end{align}
where
\begin{align}\label{e_ASK3}
R_i' = \frac{1}{n}\sum_{\nu : \nu \equiv i \mod m} H(F_\nu \mid
F_{[1, \nu -1]}) + \frac{1}{n}H\left(U_i, X_i^n \mid \mathbf{F},
K, U_{[1, i-1]}, X^n_{[1, i-1]}\right) - H(U_i).
\end{align}
Consider $B \nsubseteq \cM$, $\cA' \nsubseteq B$. For $j \in \cA'
\cap B^c$, we have
\begin{align}\nonumber
\frac{1}{n}H\left(U_B\right) + \frac{1}{n}H\left(X_B \mid X^n_{B^c}, Z_j^n\right) &= \frac{1}{n}H\left(U_B, X_B^n \mid U_{B^c}, X^n_{B^c}, Z_j^n\right)\\\nonumber &=
\frac{1}{n}H\left(F_1, ..., F_{rm}, K, U_B, X^n_B \mid U_{B^c}, X^n_{B^c},
Z_j^n\right).
\end{align}
Furthermore, since $K$ is $\ep_n$-recoverable from $(\mathbf{F},
U_{B^c}, X_{B^c}^n, Z_j^n)$ and $H(F_\nu \mid U_{B^c}, X_{B^c}^n)
= 0$ for $\nu \equiv i \mod m$ with $i \in B^c$,
\begin{align}\nonumber &\frac{1}{n}H\left(F_1, ..., F_{rm}, K,
U_B, X^n_B \mid U_{B^c}, X^n_{B^c}, Z_j^n\right)\\\nonumber &=
\frac{1}{n}\sum_{\nu =1}^{rm}H\left(F_\nu \mid F_{[1, \nu -1]},
U_{B^c}, X^n_{B^c}, Z_j^n\right) + \frac{1}{n}H\left(K\mid U_{B^c}, X^n_{B^c},
Z_j^n, \mathbf{F}\right) \\\nonumber &\quad +\frac{1}{n}\sum_{i
\in B} H\left(U_i, X_i^n \mid  U_{B^c \cap [i+1,m]}, X^n_{B^c \cap [i+1, m]}, Z_j^n,
\mathbf{F}, K, U_{[1, i-1]}, X^n_{[1, i -1]}\right) \\\nonumber &\leq
\frac{1}{n}\sum_{i \in B}\left[ \sum_{\nu : \nu \equiv i \mod m}
H\left(F_\nu \mid F_{[1, \nu -1]}\right) + H\left(U_i, X_i^n \mid
\mathbf{F}, K, U_{[1, i-1]}, X^n_{[1, i -1]}\right)\right] +\frac{\ep_n \log
|\cK| + 1}{n} \\\label{e_ASK4} &\leq  \sum_{i \in B} R_i + H(U_B),
\end{align}
where
\begin{align}\nonumber
R_i \triangleq \left(R_i' + \frac{\ep_n \log |\cK| +
1}{n}\right), \quad i \in \cM.
\end{align}
It follows from (\ref{e_ASK1}) and (\ref{e_ASK2})-(\ref{e_ASK4})  that
\begin{align}\label{e_ASK5}
\frac{1}{n}H(K) \leq H\left(\mcf\left((X_\cM, Z_i)_{i \in
\cA'}\right)\right) - \sum_{i = 1}^mR_i + \left(\ep_n + \delta_n +
\frac{\ep_n \log |\cK| + 1}{n} + \sqrt{\ep_n}\log \left(|\cX_\cM||\cZ_\cM|\right)\right),
\end{align}
where $R_\cM \in \cR\left(\cA', Z_{\cA'}\right)$ from
(\ref{e_ASK4}), and therefore
\begin{align}\label{e_ASK6}
\sum_{i = 1}^mR_i \geq R_{CO}\left(\cA', Z_{\cA'}\right).
\end{align}
Then, (\ref{e_ASK5}), (\ref{e_ASK6}) imply
\begin{align}\nonumber
\frac{1}{n}H(K) \leq C\left(\cA', Z_{\cA'}\right) + \left(\ep_n +
\delta_n + \frac{\ep_n \log |\cK| + 1}{n} +  \sqrt{\ep_n}\log
\left(|\cX_\cM||\cZ_\cM|\right)\right).
\end{align}
The proof is completed using the second part of (\ref{e_sin''}) directly, or the second part of
(\ref{e_sin'}) in the manner of \cite[Theorem 3]{CsiNar04}. This completes the converse part.
 \qed

\section*{Appendix C}\label{a_C}
\setcounter{equation}{0}
\renewcommand{\theequation}{C\arabic{equation}}
\setcounter{theorem}{0}
\renewcommand{\thetheorem}{C\arabic{theorem}}

Our proof of achievability in Theorem \ref{t_CASK} and sufficiency
in Theorem \ref{t_GSC} rely on a ``balanced coloring lemma" in
\cite{AhlCsi93}; we state below a version of it from
\cite{CsiNar04}.
\begin{lemma}\label{l_B0}\cite[Lemma 3.1]{AhlCsi93}
Let $\cP$ be any family of $N$ pmfs on a finite set $\cU$, and let
$d>0$ be such that $P \in \cP$ satisfies
\begin{align}\label{e_boundB0}
P\left(\left\{u : P(u) > \frac{1}{d}\right\}\right) \leq \ep,
\end{align}
for some $0 < \ep < (1/ 9)$. Then the probability that a randomly
selected mapping $\phi: \cU \rightarrow \{1, ..., r\}$ fails to
satisfy
\begin{align}\nonumber
\sum_{i = 1}^r \left|\sum_{u : \phi(u) = i} P(u) -
\frac{1}{r}\right| < 3\ep,
\end{align}
simultaneously for each $P \in \cP$, is less than $2Nr\exp \left(-
\frac{\ep^2d}{3r}\right)$.
\end{lemma}
In contrast to the application of Lemma \ref{l_B0} in \cite[Lemma
B.2]{CsiNar04}, our mentioned proofs call for a balanced coloring
of a set corresponding to a rv that differs from another rv for
which probability bounds are used. However, both rvs agree with
high probability when conditioned on a set of interest.

Consider rvs $U, U', V$ with values in finite sets $\cU, \cU',
\cV$, respectively, where $U'$ is a function of $U$, and a mapping
$h : \cU \rightarrow \{1, \ldots, r'\}$. For $\lambda > 0$, let
$\cU_0$ be a subset of $\cU$
such that\\
(i) $\bPr{U \in \cU_0} > 1 - \lambda^2$;\\
(ii) given $U \in \cU_0, h(U) = j, U' = u', V = v,$ there exists
$u = u(u') \in \cU_0$ satisfying
\begin{align}\nonumber
\bPr{U = u \mid h(U) = j, V= v, U \in \cU_0} =& \bPr{U' = u' \mid
h(U)=j, V=v, U \in \cU_0},\\&\label{e_bound0}\hspace{5cm} 1\leq j
\leq r', v \in \cV.
\end{align}
Then the following holds.
\begin{lemma}\label{l_B} Let the rvs $U, U', V$ and the set $\cU_0$ be as above. Further, assume that
\begin{align}\label{e_bound1}
\bP{UV}{\left\{ (u,v) : \bPr{U = u\mid V =v} > \frac{1}{d}
\right\}} \leq \lambda^2.
\end{align}
Then, a randomly selected mapping $\phi: \cU' \rightarrow \{1,
\ldots, r\}$ fails to satisfy
\begin{align}
\sum_{j=1}^{r'} \sum_{v \in {\cal V}} \bPr{h(U) = j, V =v}
\sum_{i=1}^{r} \left|\sum_{u' \in \cU': \,\phi(u')= i} \bPr{U' =
u' \mid h(U) = j, V = v}- \frac{1}{r} \right| < 14 \lambda,
\label{e_bc}
\end{align}
with probability less than $2rr'|{\cal
V}|\exp\left(-\frac{c\la^3d}{rr'}\right)$ for a constant $c > 0 $.
\end{lemma}
\textbf{Proof:} Using the condition (i) in the definition of
$\cU_0$, the left side of (\ref{e_bc}) is bounded above by
\begin{align}\nonumber
2\lambda^2 + &\sum_{j=1}^{r'} \sum_{v \in {\cal V}} \bPr{h(U) = j,
V =v, U \in \cU_0}\\ &\hspace{4cm} \sum_{i=1}^{r} \left|\sum_{u'
\in \cU': \phi(u')= i} \bPr{U' = u' \mid h(U) = j, V = v, U \in
\cU_0}- \frac{1}{r} \right|. \nonumber
\end{align}
Therefore, it is sufficient to prove that
\begin{align}\nonumber
&\sum_{j=1}^{r'} \sum_{v \in {\cal V}} \bPr{h(U) = j, V =v, U \in
\cU_0}\\ &\hspace{4cm} \sum_{i=1}^{r} \left|\sum_{u' \in \cU':
\phi(u')= i} \bPr{U' = u' \mid h(U) = j, V = v, U \in \cU_0}-
\frac{1}{r} \right| < 12\la, \label{e_bcii}
\end{align}
with probability greater than $1 - 2rr'|{\cal
V}|\exp\left(-\frac{c\la^3d}{rr'}\right)$ for a constant $c > 0$.
\\Let $q = \bP{V}{\left\{v \in \cV : \bPr{U \in
\cU_0| V = v} < \frac{1- \la^2}{3}\right\}}$. Then, since
\begin{align}\nonumber
1 - \la^2 \leq \bPr{U \in \cU_0} &\leq \sum_{v \in V :\, \bPr{U
\in \cU_0 | V = v} < \frac{1-\la^2}{3}}\hspace{-1cm} \bPr{U \in
\cU_0 | V = v}\bP{V}{v} + (1-q)\\\nonumber & <\frac{1-\la^2}{3}q +
(1-q),
\end{align}
we get from the extremities above that
\begin{align}\label{e_B1}
q < \frac{3\la^2}{2}.
\end{align}
For $u \in \cU_0$ and $v \in \cV$ satisfying
\begin{align}
\bPr{U \in \cU_0 | V = v} \geq \frac{1- \la^2}{3},\quad \bPr{U = u
| V =v, U \in \cU_0} > \frac{3}{d(1-\la^2)},
\end{align}
we have that
\begin{align}\nonumber
\bPr{U = u | V= v} > \frac{1}{d}.
\end{align}
Therefore, by (\ref{e_B1}) and (\ref{e_bound1}), it follows that
\begin{align}\nonumber
\sum_{(u,v) :\, u \in \cU_0,\, \bPr{U = u| V = v, U \in \cU_0}>
\frac{3}{d(1 - \la^2)}}\hspace{-1cm}\bPr{U =u, V =v} \leq \la^2 +
q < \frac{5\la^2}{2},
\end{align}
which is the same as
\begin{align}\nonumber
\sum_{j = 1}^{r'} \sum_{v \in \cV}& \bPr{h(U) = j, V =v, U \in
\cU_0}
\\\label{e_B2} &\sum_{u \in \cU_0 :\, \bPr{U = u| V = v, U \in \cU_0}> \frac{3}{d(1
- \la^2)}}\hspace{-1cm}\bPr{U =u| h(U)=j, V =v, U \in \cU_0} <
\frac{5\la^2}{2}.
\end{align}
The bound in (\ref{e_B2}) will now play the role of
\cite[inequality (50), p. 3059]{CsiNar04} and the remaining steps
of our proof, which are parallel to those in \cite[Lemma
B.2]{CsiNar04}, are provided here for completeness.

Setting
\begin{align}\label{e_D1}
D = \left\{ (j,v) : \sum_{u \in \cU :\, \bPr{U = u| V = v, U \in
\cU_0}> \frac{3}{d(1 - \la^2)}}\hspace{-1cm}\bPr{U =u| h(U)=j, V
=v, U \in \cU_0} \leq \frac{5\la}{2}\right\},
\end{align}
we get that
\begin{align}\label{e_D2}
\sum_{(j,v) \in D^c}\bPr{h(U) = j, V =v, U \in \cU_0} < \la.
\end{align}
Next, defining
\begin{align}\label{e_E1}
E = \left\{ (j,v) : \bPr{h(U) = j, V = v, U \in \cU_0} \geq
\frac{\la}{r'}\bPr{V =v, U \in \cU_0}\right\},
\end{align}
it holds for $(j,v) \in E$,
\begin{align}\label{e_E3}
\bPr{U =u | h(U) = j, V =v, U \in \cU_0} \leq \frac{r'}{\la}\bPr{U
=u| V =v, U \in \cU_0}.
\end{align}
Also,
\begin{align}\nonumber
\sum_{(j,v) \in E^c}\bPr{h(U) = j, V =v, U \in \cU_0} &<
\frac{\la}{r'}\sum_{j=1}^{r'}\sum_{v \in \cV}\bPr{V =v, U \in
\cU_0}\\\label{e_E2} &\leq \la.
\end{align}
Further, for $(j,v) \in E$, if
\begin{align}\label{e_E2a}
\bPr{ U = u | h(U) = j, V  = v, U \in \cU_0}> \frac{3 r'}{\la d
(1-\la^2)}
\end{align}
then from (\ref{e_E3}), we have
\begin{align}\label{e_E3a}
\bPr{U = u | V =v, U \in \cU_0} > \frac{3}{d(1 - \la^2)}.
\end{align}
Therefore, recalling the conditions that define $\cU_0$ in
(\ref{e_bound0}), we have for $(j, v) \in E\cap D$ that
\begin{align}\nonumber\displaystyle
&\sum_{ \substack{u' \in \cU' : \\\bPr{U' = u'| h(U) = j, V  = v,
U \in \cU_0}  > \frac{3 r'}{\la d (1-\la^2)}}}\hspace{-2cm}\bPr{U'
=u'| h(U)=j, V =v, U \in \cU_0}
\\\nonumber
&= \sum_{\substack{u' \in \cU' :\\ \bPr{U = u(u')| h(U) = j, V  =
v, U \in \cU_0}> \frac{3 r'}{\la d (1-\la^2)}}}\hspace{-2cm}\bPr{U
=u(u')| h(U)=j, V =v, U \in \cU_0}
\\\nonumber
& = \sum_{\substack{u \in \cU :\\ \bPr{U = u| h(U) = j, V  = v, U
\in \cU_0}> \frac{3 r'}{\la d (1-\la^2)}}}\hspace{-2cm}\bPr{U =u|
h(U)=j, V =v, U \in \cU_0}
\\\label{e_ED1}
&\leq \frac{5\la}{2},
\end{align}
where second equality is by (\ref{e_bound0}), and the previous
inequality is by (\ref{e_E2a}), (\ref{e_E3a}) and (\ref{e_D1}).
Also, using (\ref{e_D2}), (\ref{e_E2}), we get
\begin{align}\label{e_ED2}
\sum_{(j,v) \in E\cap D}\bPr{h(U) = j, V =v, U \in \cU_0} \geq 1 -
2\la.
\end{align}
Now, the left side of (\ref{e_bcii}) is bounded, using
(\ref{e_ED2}), as
\begin{align}\nonumber
&\sum_{j=1}^{r'} \sum_{v \in {\cal V}} \bPr{h(U) = j, V =v, U \in
\cU_0}
\\\nonumber &\hspace{4cm}\sum_{i=1}^{r} \left|\sum_{u' \in \cU': \phi(u')= i} \bPr{U' = u'
\mid h(U) = j, V = v, U \in \cU_0}- \frac{1}{r} \right|
\\\nonumber &\leq 4\la + \sum_{(j, v) \in E \cap D}\bPr{h(U) = j,
V =v, U \in \cU_0}\\\label{e_bc2} &\hspace{4cm} \sum_{i=1}^{r}
\left|\sum_{u' \in \cU': \phi(u')= i} \bPr{U' = u' \mid h(U) = j,
V = v, U \in \cU_0}- \frac{1}{r} \right|.
\end{align}
Using (\ref{e_ED1}), the family of pmfs $\{\bPr{U' = (\cdot) |
h(U) = j, V =v, U \in \cU_0},\quad (j,v) \in E\cap D\}$ satisfies
the hypothesis (\ref{e_boundB0}) of Lemma \ref{l_B0} with $d$
replaced by $\frac{\la(1- \la^2)d}{3r'}$ and $\ep$ replaced by
$5\la/ 2$; assume that $0 < \la < 2/45$ so as to meet the
condition following (\ref{e_boundB0}). The mentioned family
consists of at most $r'|\cV|$ pmfs. Therefore, using Lemma
\ref{l_B0},
\begin{align}\nonumber
&\sum_{j=1}^{r'} \sum_{v \in {\cal V}} \bPr{h(U) = j, V =v, U \in
\cU_0}\\\nonumber &\hspace{4cm} \sum_{i=1}^{r} \left|\sum_{u' \in
\cU': \phi(u')= i} \bPr{U' = u' \mid h(U) = j, V = v, U \in
\cU_0}- \frac{1}{r} \right| < \frac{23\la}{2}
\end{align}
with probability greater than
\begin{align}\nonumber
1 - 2 r r'|\cV|\exp\left(-\frac{25\la^3(1 -\la^2)d}{36rr'}\right)
\geq 1 - 2 r r'|\cV|\exp\left(-\frac{c\la^3d}{rr'}\right),
\end{align}
for a constant $c$. This completes the proof of (\ref{e_bcii}),
and thereby the lemma.\qed

\section*{Acknowledgements}
\noindent  The authors thank Sirin Nitinawarat for helpful
discussions.

\vspace{0.1in}

\begin{thebibliography}{10}
\providecommand{\url}[1]{#1} \csname url@samestyle\endcsname
\providecommand{\newblock}{\relax}
\providecommand{\bibinfo}[2]{#2}
\providecommand{\BIBentrySTDinterwordspacing}{\spaceskip=0pt\relax}
\providecommand{\BIBentryALTinterwordstretchfactor}{4}
\providecommand{\BIBentryALTinterwordspacing}{\spaceskip=\fontdimen2\font
plus \BIBentryALTinterwordstretchfactor\fontdimen3\font minus
  \fontdimen4\font\relax}
\providecommand{\BIBforeignlanguage}[2]{{%
\expandafter\ifx\csname l@#1\endcsname\relax
\typeout{** WARNING: IEEEtran.bst: No hyphenation pattern has been}%
\typeout{** loaded for the language `#1'. Using the pattern for}%
\typeout{** the default language instead.}%
\else \language=\csname l@#1\endcsname \fi #2}}
\providecommand{\BIBdecl}{\relax} \BIBdecl


\bibitem{AhlCsi93}
R.~Ahlswede and I.~Csisz{\'a}r, ``Common randomness in information
theory and
  cryptography--part i: Secret sharing,'' \emph{IEEE Trans. Inform. Theory},
  vol.~39, pp. 1121--1132, 1993.

\bibitem{AhlCsi07}
R.~Ahlswede and I.~Csisz{\'a}r, ``On the oblivious transfer
capacity,''
  \emph{IEEE International Symposium on Information
  Theory (ISIT)}, pp. 2061--2064, 2007.

\bibitem{CsiKor81}
I.~Csisz{\'a}r and J.~K{\"o}rner, \emph{Information theory: coding
theorems for
  discrete memoryless channels}.\hskip 1em plus 0.5em minus 0.4em\relax
  Academic Press, 1981.

\bibitem{Csi96}
I.~Csisz{\'a}r, ``Almost independence and secrecy capacity,''
\emph{Prob.
  Pered. Inform.}, vol.~32, no.~1, pp. 48--57, 1996.


\bibitem{CsiNar00}
I.~Csisz{\'a}r and P.~Narayan, ``Common randomness and secret key
generation
  with a helper,'' \emph{IEEE Trans. Inform. Theory}, vol.~46, pp. 344--366,
  March 2000.

\bibitem{CsiNar04}
I.~Csisz{\'a}r and P.~Narayan, ``Secrecy capacities for multiple
terminals,''
  \emph{IEEE Trans. Inform. Theory}, vol.~50, no.~12, pp. 3047--3061, 2004.

\bibitem{CsiNar08}
------, ``Secrecy capacities for multiterminal channel models,'' \emph{IEEE
  Trans. Inform. Theory}, vol.~54, no.~6, pp. 2437--2452, 2008.


\bibitem{GacKor73}
P.~G\'acs and J.~K\"orner, ``Common information is far less than
mutual
  information,'' \emph{Problems of Control and Information Theory}, vol.~2,
  no.~2, pp. 149--162, 1973.

\bibitem{Gal88}
R.~G. Gallager, ``Finding parity in a simple broadcast nework,''
\emph{IEEE
  Trans. Inform. Theory}, vol.~34, no.~2, pp. 176--180, 1988.


\bibitem{GriKum05}
A.~Giridhar and P.~Kumar, ``Computing and communicating functions
over sensor
  networks,'' \emph{IEEE Journ. on Select. Areas in Commun.}, vol.~23, no.~4,
  pp. 755--764, 2005.


\bibitem{KorMar79}
J.~Korner and K.~Marton, ``How to encode the modulo-two sum of
binary
  sources,'' \emph{IEEE Trans. Inform. Theory}, vol.~25, no.~2, pp. 219--221,
  1979.


\bibitem{Lov55}
M.~Loeve, \emph{Probability Theory}.\hskip 1em plus 0.5em minus
0.4em\relax Van Nostrand New York, pp. 157 and 28--42, 1955 .

\bibitem{MaIswGup09}
N.~Ma, P.~Ishwar and P.~Gupta, ``Information-theoretic bounds for
multiround
  function computation in collocated networks,'' \emph{IEEE International
  Symposium on Information Theory (ISIT)}, pp. 2306--2310, 2009.


\bibitem{Mau93}
U.~M. Maurer, ``Secret key agreement by public discussion from
common
  information,'' \emph{IEEE Trans. Inform. Theory}, vol.~39, pp. 733--742, May
  1993.

\bibitem{Mau94}
U.~M. Maurer, \emph{Communications and Cryptography: Two sides of
One
  Tapestry}, {R.E. B}lahut et al., {E}ds.~ed.\hskip 1em plus 0.5em minus
  0.4em\relax Norwell, MA: Kluwer, 1994, ch.~26, pp. 271--285.


\bibitem{NasWin06}
A.~Nascimento and A.~Winter, ``On the oblivious transfer capacity
of noisy
  correlations,'' \emph{IEEE International Symposium on Information Theory
  (ISIT)}, pp. 1871--1875, 2009.

\bibitem{OrlEl84}
A.~Orlitsky and A.~El~Gamal, ``Communication with secrecy
constraints,''
 \emph{ACM Symp. on Theory of Computing (STOC)}, pp. 217--224, 1984.


\bibitem{OrlRoc01}
A.~Orlitsky and J.~R. Roche, ``Coding for computing,'' \emph{IEEE
Trans.
  Inform. Theory}, vol.~47, no.~3, pp. 903--917, 2001.

\bibitem{Wyn74}
A.~D. Wyner, ``Recent results in the shannon theory,'' \emph{IEEE
Trans.
  Inform. Theory}, vol.~20, pp. 2--10, Jan 1974.

\bibitem{Yao79}
A.~C. Yao, ``Some complexity questions related to distributive
computing,''
  \emph{ACM Symp. on Theory of Computing (STOC)}, pp. 209--213, May 1979.

\bibitem{YeTh}
C.~Ye, ``Information theoretic generation of multiple secret
keys,'' \emph{PhD
  thesis, Dept. Elect. and Compt. Eng., University of Maryland, College Park},
  2005.

\bibitem{Ye05}
C.~Ye and P.~Narayan, ``Secret key and private key constructions
for simple
  multiterminal source models,'' \emph{Proc. Int. Symp. Inform. Theory}, pp.
  2133--2137, Sept 2005.


\end{thebibliography}


\end{document}